# Some Geophysical Aspects Of The 1908 Tunguska Event

Andrei Ol'khovatov

https://orcid.org/0000-0002-6043-9205

Independent researcher
(Retired physicist)

Russia, Moscow
email: olkhov@mail.ru

**Dedicated to the blessed memory of my grandmother ( Tuzlukova Anna Ivanovna ) and my mother ( Ol'khovatova Olga Leonidovna )**

**Abstract.** This paper is a continuation of a series of works, devoted to various aspects of the 1908 Tunguska event. In this paper its author would like to draw attention to some geophysical aspects of the 1908 Tunguska event. A review of some geophysical aspects and some minor analysis of the aspects is presented in this paper. Special attention is paid to global seismic events at those times. As the epicenter of the forest-fall was in the center of a paleovolcano, so some info concerning the paleovolcano is presented. An idea was put forward about possible relationship of the geophysical peculiarities with the atmospheric optical anomalies reported at those times.

## 1. Introduction

This paper is a continuation of a series of works in English, devoted to various aspects of the 1908 Tunguska event [Ol'khovatov, 2003; 2020a; 2020b; 2021; 2022; 2023a; 2023b; 2025a; 2025b; 2025c; 2025d]. The works can help researchers to verify the consistency of the various Tunguska interpretations with actual data. A large number of hypotheses about its causes have already been put forward. However, so far none of them has received convincing evidence.

In this paper its author (the author of this paper (i.e. A.O.) for brevity will be named as "the Author") would like to draw attention to the geophysical aspects of the Tunguska event. The Author already considered (in English) some of these aspects in [Ol'khovatov, 2003; 2020b; 2023b; 2025a] and in some others. But the Tunguska event was very complex phenomenon, so it is impossible to consider it more or less comprehensively just in several papers.

## 2. Some geophysical aspects

Let's start with the Kulikovskii paleovolcano. It was discovered in 1972 that the epicenter of the Tunguska event is in the middle of the paleovolcano later named after L. Kulik. Here is what geologists wrote [Sapronov, et al., 2001] (translated by A.O.):

"The area of the Tunguska meteorite fall is a node of deep faults of the north-western, north-eastern and submeridional directions.
<...>
Thus, geology explains many anomalous phenomena in the area of the Tunguska meteorite fall: the disturbed nature of the magnetic field, various geochemical and gas anomalies, halos of mechanical scattering of many minerals, irregularities of the radiation background, etc."

In 2005 some research about deep structure of the paleovolcano was published. Here is a fragment from [Tselomudrova et al., 2005] translated by A.O. (VIMS is here: https://vims-geo.ru , and TF is the Tunguska Phenomenon):

"In 1994, N. I. Museibov in VIMS led a study of the depth-density inhomogeneities of the Earth's crust in the northern and southwestern parts of the East Siberian Platform and the distribution patterns of ore districts, nodes, and fields with deposits of various formation types. Subsequently, using these materials, geophysicists M. I. Tselomudrova and O. V. Sidorova analyzed the depth-density inhomogeneities of the TF area and identified a deep zone of decompaction/deconsolidation (negative values of the gravity field) extending to a depth of 50 km. The decompaction/deconsolidation zone is mosaically traced from the Earth's surface to a depth of 4 km and is displayed in terms of the local components of the gravitational field (Fig. 1). Then, the decompaction/deconsolidation zone takes on an isometric shape, which is clearly visible at depths of 4-12 km (Fig. 2). Further, as it extends to a depth of 25 km, it tends to shift westward with increasing depth, with a large initial decompaction/deconsolidation zone observed at a depth of 35 km (Fig. 3).
As it can be seen from the depth-density sections shown in Figure 4, the studied volume of the Earth's crust is a tubular zone of decompaction/deconsolidation, a pipe of deep degassing and exhalation, surrounded by denser rocks."

Indeed, here is from [Alekseev et al., 2010]:

"During the expedition, hydrogen flows were measured on the routes to the Farington and Stoikovich mountains and around the Suslov crater. In some areas, the hydrogen flows related to degassing of breakage structures of the paleovolcano are anomalously high. This fact also confirms a possible endogenous origin of the geochemical anomalies (elevated concentration of microelements in the 1908 moss layers). Anomalous hydrogen flows suppress plant growth as identified in satellite photographs."

Authors of [Skublov, et al., 2011] also wrote about discovery of 2 hydrogen degassing anomalies.
The paleovolcano can be seen on a fragment of the "Kosmogeologicheskaya karta SSSR" by Ministry of Geology of the USSR, published in 1982 [Kozlovskii, 1982], which is shown on Fig.1.

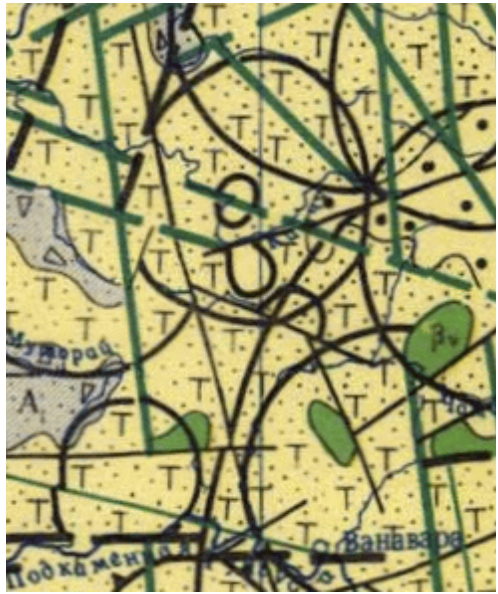

**Fig.1**

The paleovolcano is approximately in the middle of the picture. Please, pay attention to the noteworthy tectonic fault crossing the paleovolcano approximately along WNW- ESE line. Remarkably what is written in the book [Yeromenko, 1990] (translated by A.O.):

"The Berezovsko-Vanavarskii fault (BC, fig. 15, 16) within the Siberian platform was traced from the mouth of the Bakhta river to the settlements of Poligus and Mutorai and further along the route of the Tunguska meteorite in the upper reaches of the Vanavara river."

These facts allow us to look in a different way at the following results which are presented in a fragment from [Takac et al., 2024]:

"We present the first high-resolution magnetic survey of the Tunguska epicenter using unmanned aerial vehicle (UAV) magnetometry. Our survey, covering 34 km² with 100-meter profile spacing, reveals a complex pattern of magnetic anomalies, including a distinctive band aligning with the presumed impactor trajectory. While many anomalies correlate with known geological features, particularly dolerite intrusions, others suggest possible impact-related signatures. Radially averaged power spectrum analysis indicates shallow source depths (<150 m) for these anomalies. Integration of magnetic data with geological mapping and geochemical analysis of 15 rock samples shows strong correlations between iron content, magnetic susceptibility, and anomaly intensity. Notably, we observe an asymmetric pattern of magnetic features, with stronger anomalies on one side of the presumed trajectory, potentially indicating directional energy deposition during the airburst. This asymmetry aligns with recent airburst models, providing new insights into the physics of large atmospheric explosions."

Indeed, according to Olga G. Gladysheva [Gladysheva, 2022]:

"Research carried out at the epicenter showed that the Tunguska catastrophe took place at the site of an ancient volcano. However, the greatest surprise was caused by the fact that the main channel for the release of radiation energy turned out to be spatially connected with the central channel of this volcano (Gladysheva and Popov, 2016). <…> If at an altitude of ~2 km above the paleovolcano the outlines of the luminous region are rather vague, then at an altitude of ~6 km the region of maximum luminescence is located exactly above the central vent of the volcano. This is unlikely to be a coincidence."

The Author would like to draw attention to the work [Kletetschka et al., 2019] where its authors wrote: "Paleomagnetic data revealed presence of plasma during the Tunguska Event near rock surfaces". In the later article [Kletetschka et al., 2025] its authors present even more arguments about the plasma, but in the opinion of the Author, their proposed mechanism of the plasma generation does not conform with the known Tunguska facts [Ol'khovatov, 2025b].

In the opinion of the Author the data points to action of electromagnetic processes as it was briefly discussed in [Ol'khovatov, 2023b]. Anyway this aspect of the Tunguska event probably deserves a special detailed consideration.

In [Ol'khovatov, 2003] some evidences of activization of tectonic processes in the region of the Tunguska event manifestations in late June - early July 1908 were presented. The activization manifested itself, among other things, through earthquakes.

The situation at the global level was much more complex. At first, the year 1908 was the year with very low level of seismic energy release [Gutenberg and Richter, 1954]. The most comprehensive catalog of earthquakes of 1908 is [Sieberg, 1917] (however this catalog states that some of the data, especially of the end of 1908, was not received). It is important to note that this catalog is based on macroseismic data, i.e. based on reports from witnesses of seismic events. Therefore, the interpretation of its data is more complicated than the data of a catalog based on the results of instrumental registrations. However, the instrumental data is often missing for the year 1908.

Fig.2 shows the distribution of earthquakes per month in 1908, taken from Table 1 in the third part of [Sieberg, 1917]. The horizontal axis shows the month number in 1908, and the vertical axis shows the number of earthquakes reported.

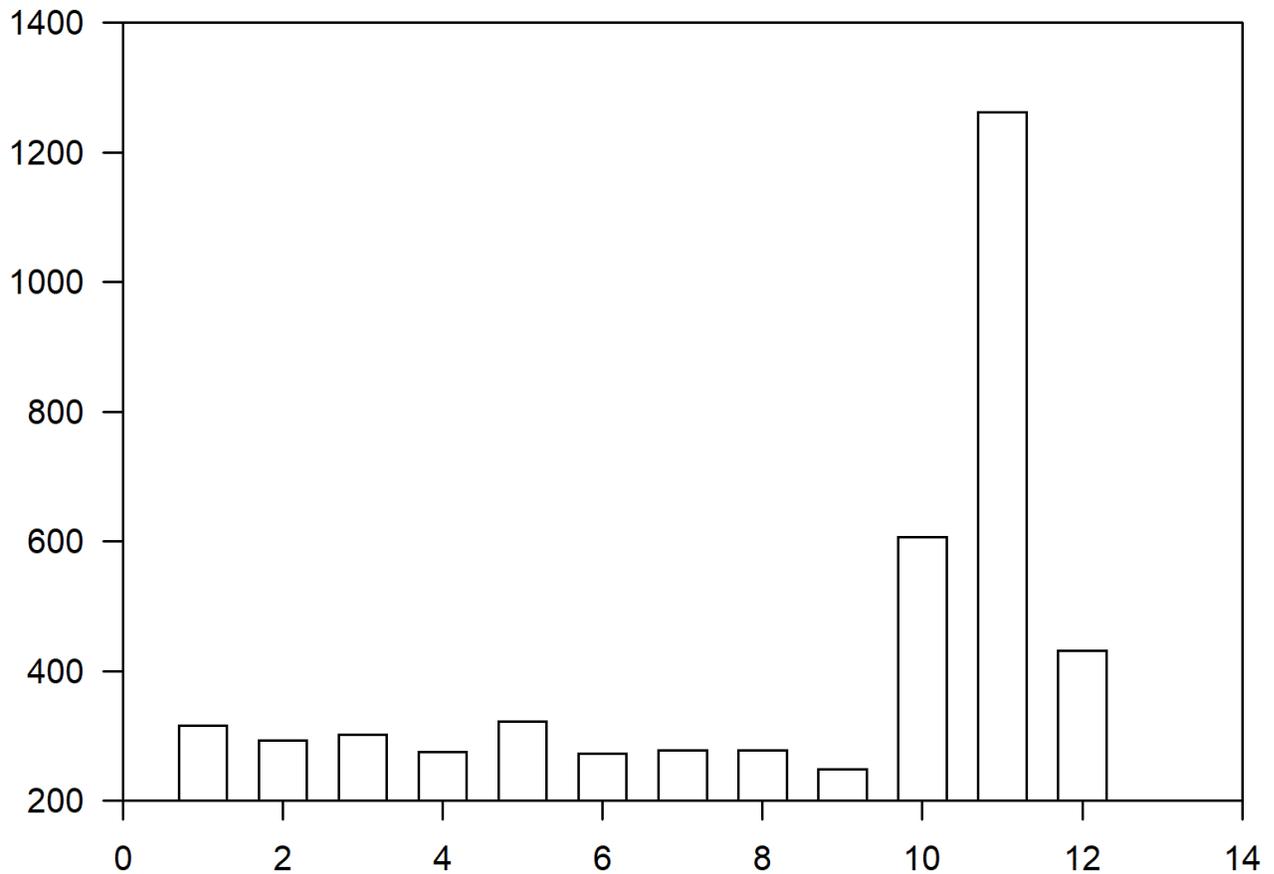

**Fig.2**

It can be seen from Fig.2 that in average during June-July 1908 the global seismic activity was on a low level.

Now let's look at the global seismic activity around June 30, 1908. Searching for all (global) earthquakes in the reviewed Bulletin of the International Seismological Centre ( https://doi.org/10.31905/D808B830 ) shows a lacuna in earthquakes between May 17 and August 20 of 1908. This does not mean that there were no earthquakes, but rather that only relatively strong earthquakes are included in this catalog.

Fortunately there is a catalog [Szirtes, 1913] with microseismic data based on registrations of seismic stations (however this catalog states that some of the data, especially of the end of 1908, was not received).  A graphic based on [Szirtes, 1913] is presented on Fig.3, where the vertical axis shows a number of earthquakes reported all over the Earth per day. The horizontal axis shows days from June 15 to July 15, 1908. The data for June 30, and July 1 is marked with a red color.

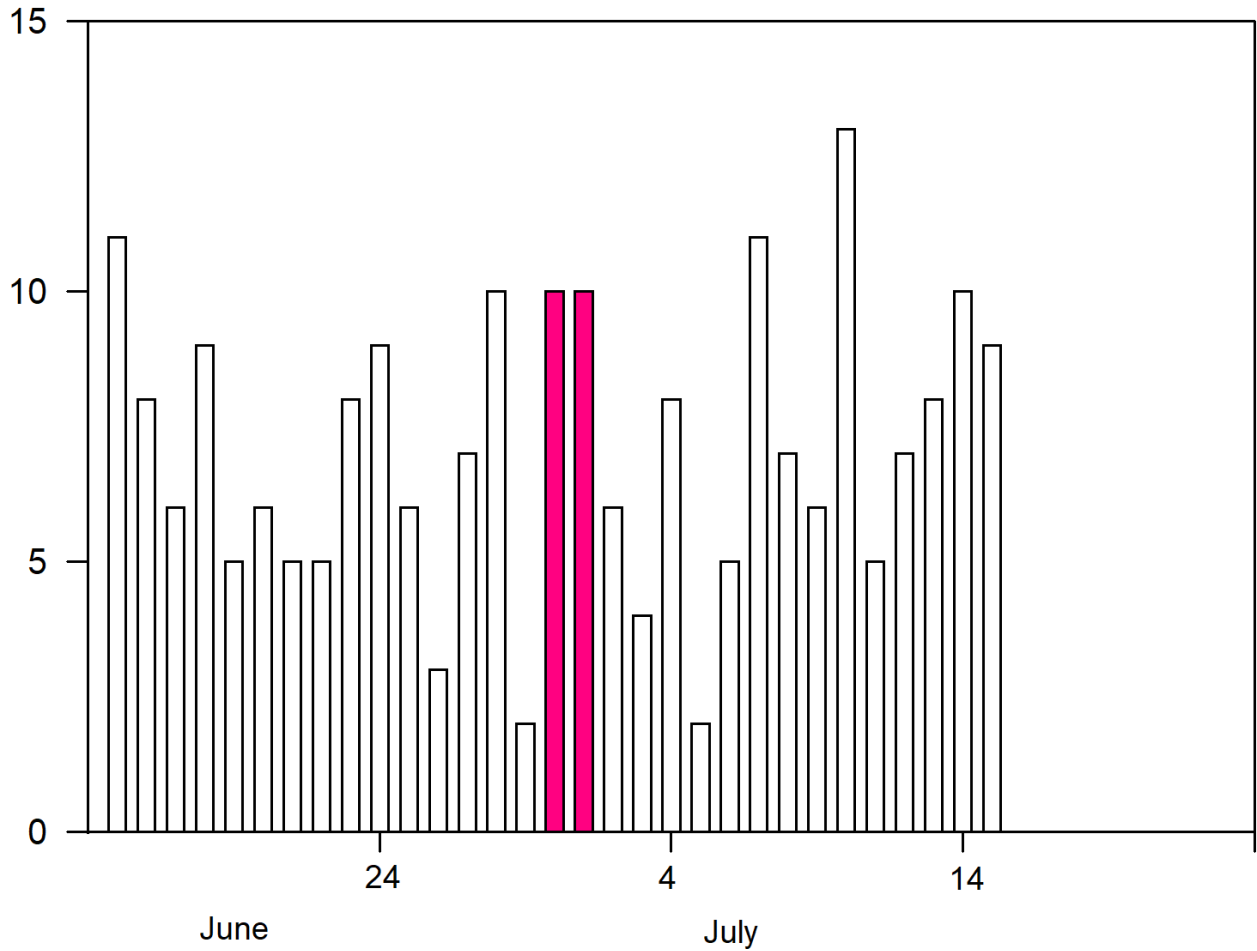

**Fig.3**

From Fig.3, it can be seen that perhaps the only small peculiarity is the presence of two moderate-amplitude equal upsurges following each other. This is the only occurrence in a whole month.

Let's look at the data from [Sieberg, 1917] which has a lot of macroseismic data. In [Ol'khovatov, 2003] there was already a graphic, but later the Author discovered some minor inaccuracies in the plotting, which however did not affect the main point - some upsurge of the global seismic activity on June 30 and July 1, 1908. Anyway the Author made a new graphic and on an extended time interval based on [Sieberg, 1917]. It is presented on Fig.4, where the vertical axis shows a number of earthquakes reported all over the Earth per day according to their universal times. The horizontal axis shows days from June 20 to July 10, 1908.

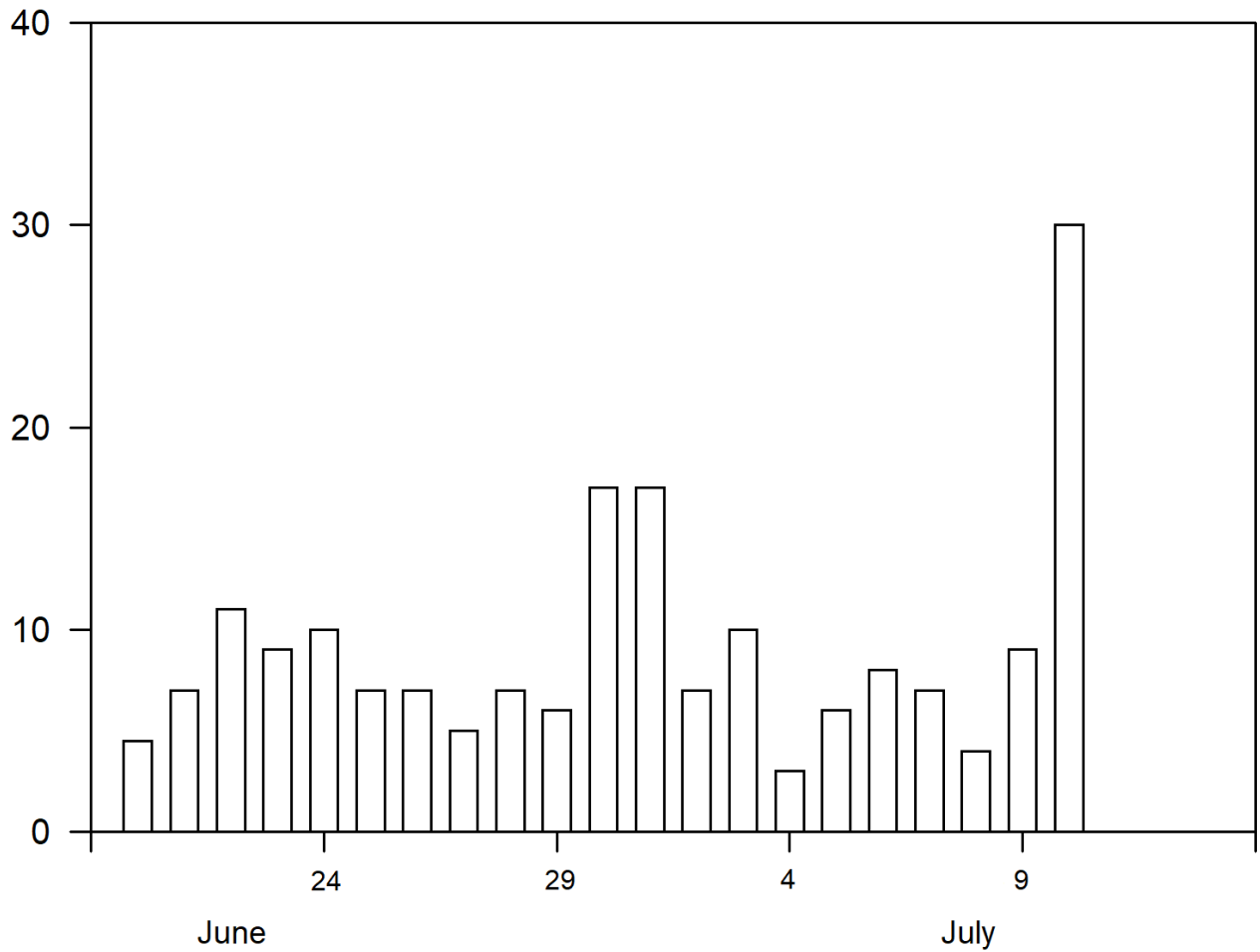

**Fig.4**

The value for June 20 was taken as 4.5 because in one place the date of the earthquake was given in [Sieberg, 1917] as of June 20 or 19. The upsurge on June 30 and July 1 is clearly seen. Interestingly, the amplitude of both peaks (of June 30 and July 1) turned out to be the equal to each other on both Figures. Also on July 10 (on Fig.4) there is even larger upsurge. Remarkably the upsurge on July 10 was caused mainly by a series of earthquakes in one region, while the upsurge on June 30 - July 1 was caused by earthquakes in several regions of the Earth. Comparing Fig.3 and Fig.4 hints that on June 30 and July 1, 1908 there was some moderate increase of the number of small and moderate earthquakes, a part of which was not registered by seismic stations.

Thus it looks like there were some peculiarities in the global seismicity.

## 3. Discussion

In the previous works, the Author has already presented various geophysical peculiarities. This publication adds some details to them.

Now the Author would like to propose a possible mechanism which could make some input into atmospheric optical anomalies of late June - early July, 1908. The mechanism is associated with the endogenic hydrogen degassing with following conversion (with oxygen) into water vapor in the upper atmosphere. Let's explain.

As mentioned earlier, hydrogen degassing has been detected in the vicinity of the Tunguska epicenter. Here are some words about the hydrogen degassing in general [Zgonnik, 2020]:

"These same studies showed that hydrogen concentrations varied greatly in space and time: it was reported that even in 50 cm-spaced monitoring holes, $H_2$ concentrations could be 3 and 7000 ppm simultaneously. Even more surprising, monitoring at several locations situated up to 50 km apart showed $H_2$ concentrations increasing simultaneously but independently of local meteorological conditions (Satake et al., 1985), suggesting a global-scale process as being the cause of this synchronous behavior."

It should be noted that there are several other paleovolcanic structures in the region in addition to the Kulikovskii paleovolcano. The region of the Tunguska event manifestations is closely adjacent to the Baikal Rift Zone. In these regions, tectonic processes became more active at the time in question [Ol'khovatov, 2003].

The possible intensity of hydrogen degassing in the region of interest can be estimated from the following data related to one of the locations in Lake Baikal [Isaev and Pastukhov, 2022] translated by A.O.:

"For example, an area of 5-6 $m^2$ of water surface releases approximately 20.0 $m^3$ of hydrogen per hour (with an average hydrogen content of 40% by volume), which amounts to 480 $m^3$ per day."

Hydrogen is a light gas, so it will rise. The prevailing winds at high altitudes will carry the hydrogen westward, and at least a part of it will gradually turn into water vapor as it can bond with oxygen via several ways. Of course this will also affect other chemical reactions too, but they are beyond the scope of this paper. The formation of water vapor could make it easier to interpret the observed atmospheric optical anomalies.

In the Author's opinion similar planetary degassing should be taken into account while calculating the planetary climate variations, but this aspect is beyond the scope of this paper.

## 4. Conclusion

The general conclusion is that the Tunguska event was a very complex phenomenon, with probably large-scale (global?) processes involved. In some of his other works the Author also considers a possibility of some other factors role (solar activity, etc.). Research of the Tunguska event requires the participation of experts in various fields.

## ACKNOWLEDGEMENTS

The author wants to thank the many people who helped him to work on this paper, and special gratitude to his mother - Ol'khovatova Olga Leonidovna (unfortunately she didn't live long enough to see this paper published...), without her moral and other diverse support this paper would hardly have been written.